\documentclass[conference]{IEEEtran}

\ifCLASSINFOpdf
\else
\fi

\usepackage{amsbsy,amscd,amsfonts,amsgen,amsmath,amsopn,amssymb,amstext,amsthm,amsxtra}
\usepackage{graphicx}
\usepackage{verbatim}
\usepackage{citesort}
\usepackage{url}
\usepackage{color}

\begin{document}
\title{Performance Regions in Compressed Sensing\\ from Noisy Measurements}

\author{\IEEEauthorblockN{Junan Zhu and Dror Baron}
\IEEEauthorblockA{Department of Electrical and Computer Engineering\\North Carolina State University; Raleigh, NC 27695, USA\\Email: \{jzhu9,barondror\}@ncsu.edu}}

%
\maketitle

\begin{abstract}
In this paper, compressed sensing with noisy measurements is addressed. The theoretically optimal reconstruction error is studied by evaluating Tanaka's equation. The main contribution is to show that in several regions, which have different measurement rates and noise levels, the reconstruction error behaves differently. This paper also evaluates the performance of the belief propagation (BP) signal reconstruction method in the regions discovered. When the measurement rate and the noise level lie in a certain region, BP is suboptimal with respect to Tanaka's equation, and it may be possible to develop reconstruction algorithms with lower error in that region.
\end{abstract}



\IEEEpeerreviewmaketitle

\section{Introduction}
\subsection{Motivation}
Compressed sensing (CS) is an emerging area of signal processing that allows to reconstruct sparse signals from a reduced number of measurements \cite{DonohoCS,CandesRUP,BaraniukCS2007}. Because real world applications involve noisy measurements, CS with noise has drawn a lot of attention \cite{HN05,Sarvotham06,Wainwright2007,Akcakaya2010}.

In recent work, Wu and Verd{\'u} \cite{Wu2011} defined the \textit{noise sensitivity} as the ratio of the reconstruction error to the noise variance. The reconstruction is called \textit{robust} if the noise sensitivity is finite. Wu and Verd{\'u} proved that the reconstruction is robust only when the measurement rate $R$, which is the ratio of the number of measurements to the signal length, exceeds a certain threshold. Wu and Verd{\'u}'s threshold deals with the case where the noise is low. Unfortunately, the measurement noise in real-world applications might be greater than that required by Wu and Verd{\'u}. Thus, it is of interest to investigate the behavior of the CS reconstruction error in regions where there is more noise.

\subsection{Contribution}
Tanaka's fixed point equation provides the fundamental information theoretical limit on the \emph{reconstruction performance}, which is quantified as the minimum mean square error (MMSE) of signal reconstruction in the presence of measurement noise \cite{Tanaka2002,GuoVerdu2005,GuoWang2008,GuoBaronShamai2009,GuoTanaka2009}. In this paper, we use Tanaka's equation to evaluate the reconstruction performance for sparse Gaussian signals. That is, each element  follows a Gaussian distribution with probability $p$, while it is zero with probability $1-p$. We call $p$ the \emph{sparsity rate} of the signal.

\begin{figure}[t]
\centering
\includegraphics[width=9cm]{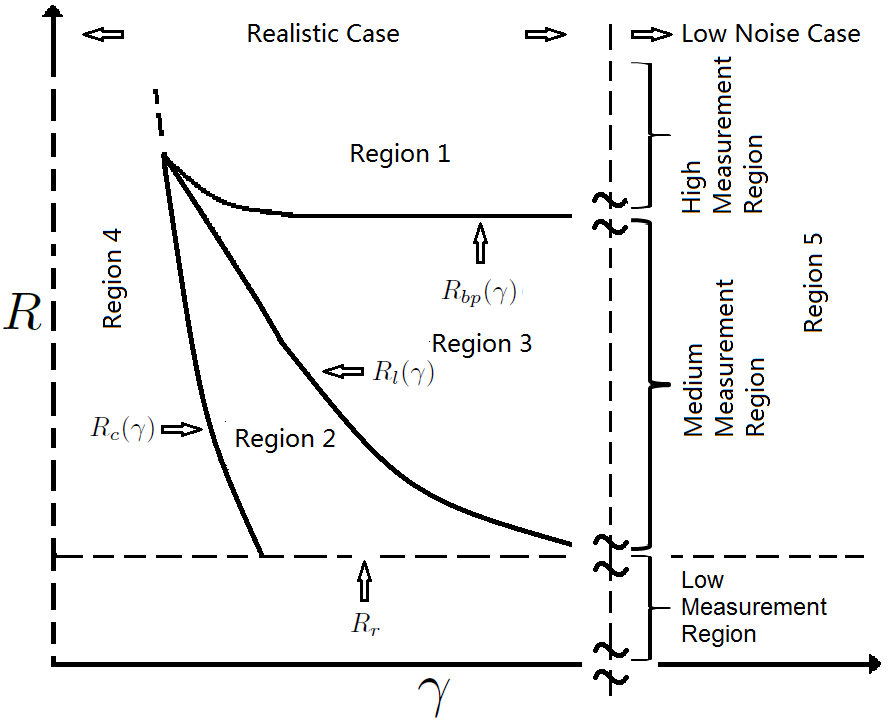}
\caption{Different regions and thresholds in CS reconstruction.}\label{fig.RegionSchematics}
\end{figure}

It might seem that lower noise always results in better reconstruction performance. However, the main result of this paper is to show that \emph{the behavior of the reconstruction performance is more nuanced in CS}. There are several different performance regions and thresholds that separate these regions, as illustrated in Figure~\ref{fig.RegionSchematics}. (See Section III for a detailed discussion. Note that the $\gamma$ in Figure~ \ref{fig.RegionSchematics} stands for the inverse noise level; larger $\gamma$ means less noise.)
\begin{itemize}
  \item \textbf{Low measurement region:} The measurement rate $R$ is too small. Robust reconstruction is impossible.
  \item \textbf{High measurement region:} $R$ is sufficiently large. This region consists of Region~1, and parts of Regions~4 and~5. Increasing the inverse noise level $\gamma$ leads to an immediate improvement in performance.
  \item \textbf{Medium measurement region:} $R$ is modest. This region consists of Regions~2 and 3, and parts of Regions~4 and 5. In the following we add further detail about the medium measurement region.
  \begin{enumerate}
    \item \textbf{Region~4:} (Including the portion of Region~4 in the high measurement region.) The noise is high in this region, and thus, the reconstruction quality is poor.
    \item \textbf{Regions~2 and 3:} The noise is modest in these regions. For fixed $R$, when the noise decreases ($\gamma$ increases) we leave Region~4 and enter Region~2. The performance hits a barrier and stays roughly constant. This barrier is broken through when we further increase $\gamma$ and enter Region~3, and the performance improves significantly.
    \item \textbf{Region~5:} (Including the portion of Region~5 in the high measurement region.) The noise is low in this region. As the noise decreases to zero, the reconstruction error vanishes. This is the case that Wu and Verd{\'u} studied \cite{Wu2011}.
  \end{enumerate}
\end{itemize}
We call the medium and high measurement regions the \emph{robust region}, because the noise sensitivity within these regions is finite~\cite{Wu2011}; further insights are given in Section III.B. Of particular interest within the robust region is its portion where the noise is medium to high (Regions~1-4); we call this the \emph{realistic case}.

In this paper, we also discuss belief propagation (BP) \cite{Guo2006,GuoWang2008,Rangan2010,CSBP2010,RanganGAMP2010}, which is a signal reconstruction method that often achieves the theoretically optimal performance of CS in the large system limit. Having discussed the regions mentioned above, we evaluate the performance of BP in Section IV. \emph{Another key finding of this paper is that BP is suboptimal in Region~3.}

\subsection{Organization}
The rest of this paper is organized as follows. Section II discusses background material including Tanaka's fixed point equation. In Section III, we describe the problem setting and explicitly explain the different performance regions in detail. We then compare the performance of BP to the theoretically optimal performance of CS in Section IV. Section V concludes.

\section{Background}
We begin by defining some notations. Let $x$ be the input signal, with each individual element being denoted by $x_i$, $i\in \{1,2,...,N\}$. The input $x$ is acquired by multiplying it by a measurement matrix $\Phi\in\mathbb{R}^{M\times N}$, which has unit norm rows. Our model also involves additive measurement noise, which is denoted by $z$. The input signal $x$, the channel noise $z$, and the entries of $\Phi$ are independent and identically distributed (i.i.d.), respectively. The measurement vector $y\in \mathbb{R}^M$ is given by
\begin{equation}\label{eq.vector}
    y=\sqrt{\gamma}\Phi x+z,
\end{equation}
where $\gamma$ is the amplification of the channel, which represents the inverse noise level \eqref{eq.vector} and thus relates to the signal-to-noise ratio (SNR). The number of measurements, which is also the length of $y$, is denoted by $M$. Finally, we define the \emph{measurement rate} as
\begin{align*}
    R=\frac{M}{N}.
\end{align*}

Guo and Wang \cite{GuoWang2008} stated that the fundamental information theoretical characterization of the optimal reconstruction performance in CS remains the same when using a fixed $R$, where $M$ and $N$ scale to infinity, i.e., $\lim_{N\rightarrow\infty}\frac{M}{N}=R$; this is called the large system limit. We discuss signal reconstruction performance as a function of the measurement rate $R$ and the inverse noise level $\gamma$ in the large system limit.

According to Guo and Wang \cite{GuoWang2008}, the problem of estimating each scalar $x_i$ given the vector $y$ in the vector channel \eqref{eq.vector} is asymptotically equivalent to estimating the same symbol through a scalar channel,
\begin{equation}\label{eq.scalar}
    \widetilde{y}_i=\sqrt{\eta\gamma R}x_i+\widetilde{z}_i,
\end{equation}
where $\widetilde{y}_i$ and $\widetilde{z}_i$, $i\in\{1,2,...,N\}$, are the sufficient statistics for $y$ and $z$ in the vector channel \eqref{eq.vector}, respectively, and $\eta\in (0,1)$ is the {\em degradation} of the scalar channel relative to the original vector channel. The amount of noise in \eqref{eq.scalar} can be computed by finding a fixed point for $\eta$ in the following equation by Tanaka \cite{Tanaka2002,Wu2011},
\begin{equation}\label{eq.TE}
    \frac{1}{\eta}=1+\gamma\cdot \mbox{MMSE},
\end{equation}
where MMSE is the minimum mean square error of the estimation in the equivalent scalar channel \eqref{eq.scalar}. Tanaka's equation may give more than one fixed point (an example for multiple fixed points can be found in \cite{Wu2011}); the correct fixed point minimizes the free energy \cite{GuoBaronShamai2009,GuoTanaka2009},
\begin{equation}\label{eq.FreeEnergy}
    E(\eta)=I(x_i;\sqrt{\eta\gamma R}x_i+\widetilde{z}_i)+\frac{R}{2}(\eta-1-\log_{2} (\eta)),
\end{equation}
where $I(\cdot;\cdot)$ is the mutual information \cite{Cover06}.
Note that Tanaka's results are based on the replica method in statistical physics, and are not rigorous \cite{GuoVerdu2005}. On the other hand, replica analyses in compressed sensing can be made rigorous in some cases; see Bayati and Montanari \cite{Bayati2011} for example.

By evaluating \eqref{eq.TE} and \eqref{eq.FreeEnergy}, we can compute the degradation $\eta$ of the scalar channel, as well as the MMSE. With this MMSE, we can determine whether a certain measurement rate $R$ provides acceptable reconstruction performance. Note that Tanaka's equation can also yield information theoretical limits on the lowest attainable mean error when non-quadratic error metrics are considered \cite{Tan2012signal}.


\section{Reconstruction in the robust region}
Having discussed Tanaka's equation, we now employ it in analyzing the sparse Gaussian case, with the i.i.d. input signal $x$ following the distribution function,
\begin{equation}\label{eq.sparseG}
    f_{X_i}(x_i)=p\cdot\frac {1}{\sqrt{2\pi}}e^{-\frac{x_i^2}{2}}+(1-p)\cdot\delta_0(x_i)
\end{equation}
for $i\in\{1,...,N\}$, where $\delta_0(x)$ denotes the delta function with probability mass at the origin. We further assume that the noise $z$ is i.i.d. unit variance zero mean Gaussian distributed,
\begin{equation}\label{eq.noise}
f_{Z_i}(z_i)=\frac {1}{\sqrt{2\pi}}e^{-\frac{z_i^2}{2}},\quad i\in\{1,...,N\}.
\end{equation}
In the sparse Gaussian case, SNR$=\gamma\cdot p$ for the vector channel~\eqref{eq.vector}, whereas SNR$=\eta\gamma p$ for the scalar channel~\eqref{eq.scalar}.

Tanaka's equation \eqref{eq.TE} allows us to investigate the vector problem~\eqref{eq.vector} by analyzing the corresponding scalar channel~\eqref{eq.scalar}. It can be shown that in the scalar channel~\eqref{eq.scalar}, the MMSE of the sparse Gaussian case satisfies
\begin{equation}\label{eq.MMSE}
\begin{split}
   &\mbox{MMSE}=p-\frac{p^2a R}{\sqrt{2\pi}(a R+1)^{2.5}}\times\\
   &\int_y \frac{y^2}{p\cdot e^{\frac{1}{2(a R+1)}y^2}+(1-p)\sqrt{a R+1}\cdot e^{\frac{1-a R}{2(a R+1)}y^2}}dy,
\end{split}
\end{equation}
where $a=\eta\gamma$ for notational simplicity.

After specifying the sparsity rate $p$, the inverse noise level $\gamma$, and the measurement rate $R$, we obtain the channel degradation $\eta$ and thus the MMSE in \eqref{eq.TE}. In the following two subsections, we first discuss the reconstruction performance in the low noise case, i.e., Region~5, and then discuss the realistic case, which includes Regions~1-4.

\subsection{CS reconstruction performance in the low noise case}
Wu and Verd{\'u} \cite{Wu2011} analyzed the robustness of reconstruction performance (MMSE) in the low noise region (Region~5) where $\gamma\rightarrow \infty$,
\begin{equation}\label{eq.approxMMSE}
    \mbox{MMSE}(p,\eta\gamma R)=\frac{p}{\eta \gamma R} (1+o(1)),
\end{equation}
where the $o(1)$ term vanishes as $\gamma\rightarrow\infty$. Substituting \eqref{eq.approxMMSE} into \eqref{eq.TE} and solving for $\eta$, we obtain
\begin{equation*}
\eta=1-\frac{p}{R} (1+o(1))
\end{equation*}
and
\begin{equation*}
\mbox{MMSE}(p,\eta\gamma R)=\frac{p}{\gamma(R-p)}(1+o(1)).
\end{equation*}

Wu and Verd{\'u} \cite{Wu2011} define the noise sensitivity as MMSE divided by the noise variance. In this case the noise sensitivity
\begin{equation*}
    \frac{\mbox{MMSE}(p,\eta\gamma R)}{1/\gamma}=\frac{p}{R-p}(1+o(1))
\end{equation*}
is finite only when $R>p$. Therefore, the threshold for robust reconstruction is $R=p$ for sparse Gaussian signals.

As the reconstruction is not robust below the threshold, i.e., $R<p$, we are only interested in the robust region, $R>p$. Also, considering that the measurement rate in a CS system is lower than 1, we concentrate on the case when $p<R<1$. We can see that the MMSE \eqref{eq.MMSE} is a function of $a=\eta\gamma$. If $\gamma$ grows and $a$ stays constant for a fixed $R$, then we will obtain a constant MMSE and an $\eta$ that decreases to zero. As a result, \eqref{eq.TE} can be rewritten as
\begin{equation}\label{eq.rewriteTE}
    1-a\cdot \mbox{MMSE}(p,a R)=\eta\xrightarrow{\gamma\rightarrow\infty} 0.
\end{equation}
That is, the fixed point of the form $\eta=\frac{a}{\gamma}$ satisfies \eqref{eq.TE}. Therefore, when solving Tanaka's equation in Region~5, there could be several fixed points: (\emph{i.}) the first fixed point is $\eta_1=\frac{\widetilde{a}_1}{\gamma}$, where $\widetilde{a}_1$ is a constant for a fixed $R$; (\emph{ii.}) the second is $\eta_2=\frac{\widetilde{a}_2}{\gamma}$, where $\widetilde{a}_2$ is another constant, which is larger than $\widetilde{a}_1$, for a fixed $R$; and (\emph{iii.}) the third is $\eta_3=1-\frac{p}{R}$. The correct fixed point minimizes the free energy \eqref{eq.FreeEnergy} and is determined later.

For $R>R_{bp}(\gamma)$, there is only one fixed point $\eta_3=1-\frac{p}{R}$ and there is no need to evaluate the free energy. At $R_{bp}(\gamma)$, there are exactly two fixed points, which are $\eta_1=\frac{\widetilde{a}_1}{\gamma}$ and $\eta_3=1-\frac{p}{R}$. When reducing $R$ below the threshold $R_{bp}(\gamma)$, $\eta_2$ appears. Note that in the realistic case with more noise, the approximations above for $\eta_1$, $\eta_2$, and $\eta_3$ become imprecise; the discussion for this case appears in Section III.B.

If the optimal performance is obtained by $\eta_1$ or $\eta_2$, then the SNR of the corresponding scalar channel remains constant as the noise declines, and the channel ends up with a roughly fixed reconstruction error. However, because $\widetilde{a}_2$ is larger than $\widetilde{a}_1$, the SNR will be greater in the case of $\eta_2$ than that of $\eta_1$. Thus, if $\eta_1=\frac{\widetilde{a}_1}{\gamma}$ ends up being the correct fixed point, then the performance will be the worst among the three fixed points. Therefore, we prefer the case that $\eta_3=1-\frac{p}{R}$ minimizes the free energy, because this fixed point $\eta_3$ leads to the lowest MMSE.

We now evaluate the free energy of each fixed point. The one that minimizes the free energy \eqref{eq.FreeEnergy} corresponds to the theoretically optimal performance of the channel. As we have already seen that in Region~5 we could have $\eta_1=\frac{\widetilde{a}_1}{\gamma}$, $\eta_2=\frac{\widetilde{a}_2}{\gamma}$, and $\eta_3=1-\frac{p}{R}$, we need to evaluate the free energy of each fixed point.

The mutual information of the channel \eqref{eq.scalar} can be approximated as in \cite{WuVerdu2011},
\begin{equation}\label{eq.approxMutualInfo}
    I(x_i;\widetilde{y}_i)\sim \frac{1}{2}p \ln (\eta \gamma R).
\end{equation}
Plugging \eqref{eq.approxMutualInfo} into \eqref{eq.FreeEnergy},
for $\eta_1=\frac{\widetilde{a}_1}{\gamma}$,
\begin{equation}\label{eq.EofEta1}
\begin{split}
    E(\eta_1)\sim &\frac{p-\frac{R}{\ln (2)}}{2} \ln (\widetilde{a}_1)+\frac{p}{2}\ln (R)+\\
    &\frac{R}{2\gamma} \widetilde{a}_1-\frac{R}{2}+\frac{R}{2}\log_2 (\gamma);
\end{split}
\end{equation}
for $\eta_2=\frac{\widetilde{a}_2}{\gamma}$,
\begin{equation}\label{eq.EofEta2}
\begin{split}
    E(\eta_2)\sim &\frac{p-\frac{R}{\ln (2)}}{2} \ln (\widetilde{a}_2)+\frac{p}{2}\ln (R)+\\
    &\frac{R}{2\gamma} \widetilde{a}_2-\frac{R}{2}+\frac{R}{2}\log_2 (\gamma);
\end{split}
\end{equation}
and for $\eta_3=1-\frac{p}{R}$,
\begin{equation}\label{eq.EofEta3}
    E(\eta_3)\sim \frac{p-\frac{R}{\ln (2)}}{2}\ln (1-\frac{p}{R})+\frac{p}{2}\ln (\frac{R}{e})+\frac{p}{2}\ln (\gamma).
\end{equation}
Considering that $p<R<1$, we have
\begin{equation*}
\begin{split}
&E(\eta_1)\sim \frac{R}{2} \log_2 (\gamma),
\quad
E(\eta_2)\sim \frac{R}{2} \log_2 (\gamma),
\quad \\
&\mbox{and} \quad
E(\eta_3)\sim \frac{p}{2} \ln (\gamma)
\end{split}
\end{equation*}
for large $\gamma$. The only difference between \eqref{eq.EofEta1} and \eqref{eq.EofEta2} is in the constants $\widetilde{a}_1$ and $\widetilde{a}_2$. Provided that $\widetilde{a}_1<\widetilde{a}_2$ and $R>p$, we obtain that $E(\eta_3)<E(\eta_1)<E(\eta_2)$. Thus, the correct $\eta$ is $\eta_3=1-\frac{p}{R}$, and MMSE$(p,\eta\gamma R)=\frac{p}{\gamma (R-p)} (1+o(1))$ decreases to zero as $\gamma$ increases to $\infty$. That is, in the robust region the reconstruction error vanishes as long as $\gamma$ is sufficiently large (Region~5).

\subsection{CS reconstruction performance in the realistic case}
The MMSE can be driven down to zero in the robust region by driving down the noise to zero. However, we cannot use an arbitrarily large $\gamma$, owing to the physical characteristics of the measurement system. Thus, it is worthy to study the realistic case (Regions~1-4) where the noise is not small.

In the realistic case, the approximation of the fixed points might be imprecise. Also, the approximation of the free energy~\eqref{eq.EofEta1} -- \eqref{eq.EofEta3} does not apply, because $\gamma$ does not go to $\infty$. Therefore, we must calculate each fixed point and the corresponding free energy~\eqref{eq.FreeEnergy}. Evaluating Tanaka's equation numerically gives Figure~\ref{fig.RegionSchematics}, which appeared in Section I. The following discussion explains the figure in detail.
\begin{enumerate}
  \item Above the \emph{BP threshold} $R_{bp}(\gamma)$, BP is advantageous (Section IV). Below $R_{bp}(\gamma)$, the MSE performance of BP is not satisfactory.
  \item The threshold that separates the robust region and the low measurement (unrobust) region is called the \emph{robust threshold}; it is denoted by $R_r$, and was characterized comprehensively by Wu and Verd{\'u} \cite{Wu2011}.
  \item In Regions~1 and 4 there is only one fixed point; these two regions merge together as $R$ is increased. The difference is that the SNR in Region~1 is higher, thus the reconstruction error in Region~1 is lower than that in Region~4.
  \item The boundary separating Region~2 and 4 is the \emph{consistency threshold} $R_c(\gamma)$, beyond which the reconstruction error stays roughly consistent for a fixed $R$ in Region~2, because the smallest $\eta_1\approx\frac{\widetilde{a}_1}{\gamma}$ minimizes the free energy, and the scalar channel SNR$=\eta\gamma p$ is roughly constant.
  \item The boundaries separating Regions~1 and 4 from Regions~2 and 3 are the \emph{consistency threshold} $R_c(\gamma)$ and the BP threshold $R_{bp}(\gamma)$; they are the thresholds bounding Regions 2 and 3, where three fixed points emerge. On $R_c(\gamma)$ and $R_{bp}(\gamma)$, there are exactly two fixed points.
  \item Regions~2 and 3 are separated by the \emph{low noise threshold} $R_l(\gamma)$, which means that above this threshold, the MMSE behaves as in the low noise case. In Region~3, which lies between the thresholds $R_l(\gamma)$ and $R_{bp}(\gamma)$, the reconstruction error decreases as $\gamma$ increases for a fixed $R$, because the biggest $\eta$ (not necessarily equal to $1-\frac{p}{R}$) minimizes the free energy.
\end{enumerate}

As we have seen, in Region~2 the MMSE is roughly fixed for a certain $R$. On the other hand, in Regions~1 and 3 the reconstruction error decreases as $\gamma$ is increased. However, the boundary between Regions~2 and~3, $R_l(\gamma)$, approaches $R_r$ as $\gamma$ increases. Therefore, for a fixed measurement rate greater than $R_r$, if we increase $\gamma$, then we will eventually traverse $R_l(\gamma)$ and obtain a significant reduction in the reconstruction error as we move from Region~2 to Region~3.

\begin{figure}[t]
\centering
\includegraphics[width=9cm]{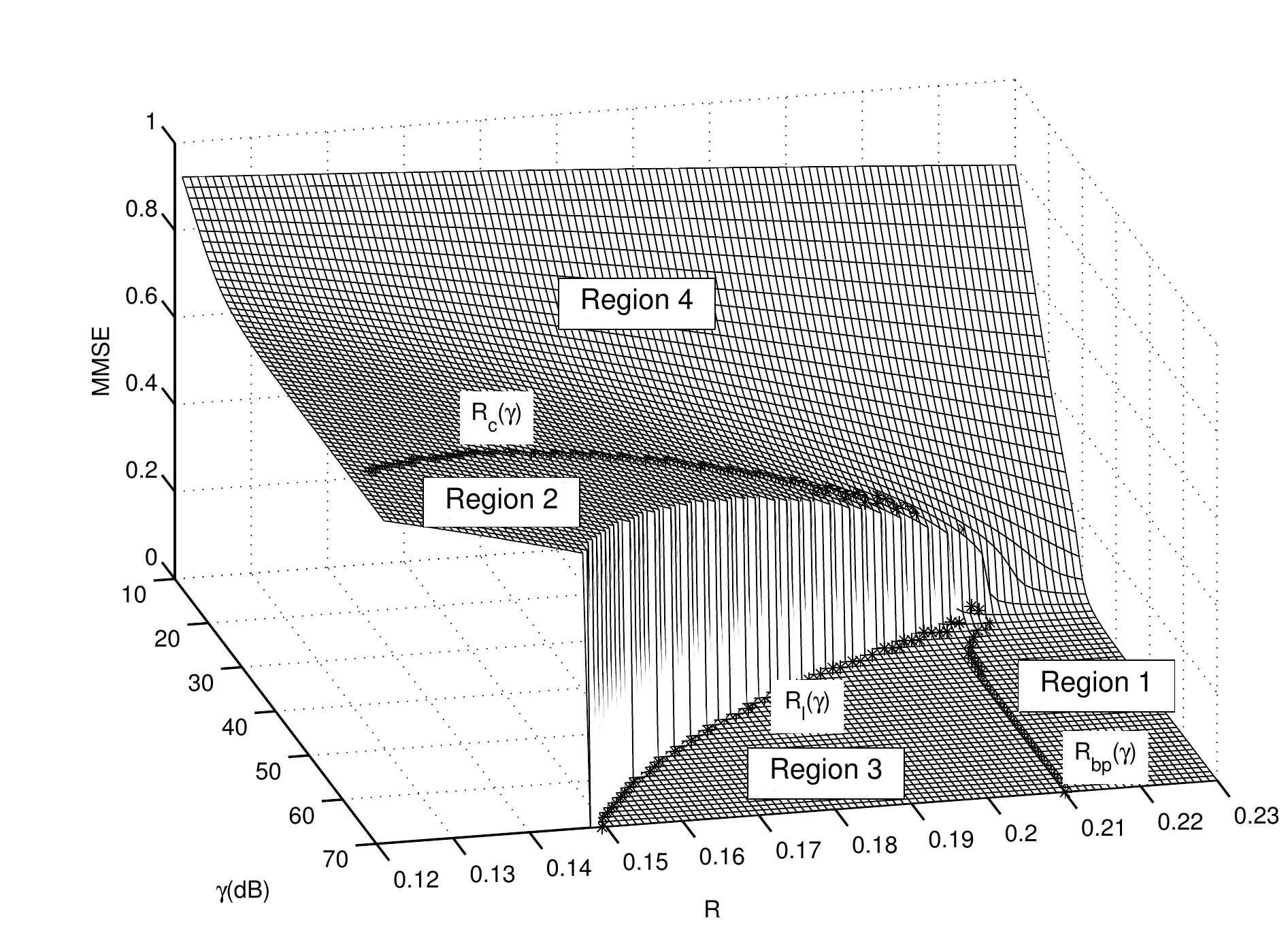}
\caption{Theoretical CS performance of a sparse Gaussian signal ($p=0.1$) provided by Tanaka's equation. Regions~1-4, as well as the thresholds $R_{bp}(\gamma)$, $R_l(\gamma)$, and $R_c(\gamma)$, are marked out.}\label{fig.SurfMMSE}
\end{figure}

By specifying the sparse Gaussian problem as $p=0.1$ and $R\in (0.12,0.23)$, we obtain the MMSE, which is shown in Figure~\ref{fig.SurfMMSE}; Regions~1-4, as well as the thresholds $R_{bp}(\gamma)$, $R_c(\gamma)$, and $R_l(\gamma)$, are also marked out in the figure.

We can see in Figure~\ref{fig.SurfMMSE} that for a relatively small $R$, when $\gamma$ increases, the MMSE first decreases slightly, then it remains roughly constant, and after a further increase in $\gamma$ the MMSE experiences a significant reduction and vanishes. On the other hand, for large $R$, increasing $\gamma$ always leads to a prominent decrease in the MMSE.

\section{Performance of Belief Propagation}
Having characterized the different performance regions, we aim in this section to shed light on the performance regions that practitioners may encounter when implementing CS reconstruction algorithms in real systems. Belief propagation (BP) \cite{Rangan2010} is known to achieve the MMSE performance of the smallest fixed point, and it is optimal for many problems of interest. Thus, it is interesting to evaluate the performance of BP on the regions that were discussed in Section III.

\begin{figure*}[t]
\centering
\includegraphics[width=14cm]{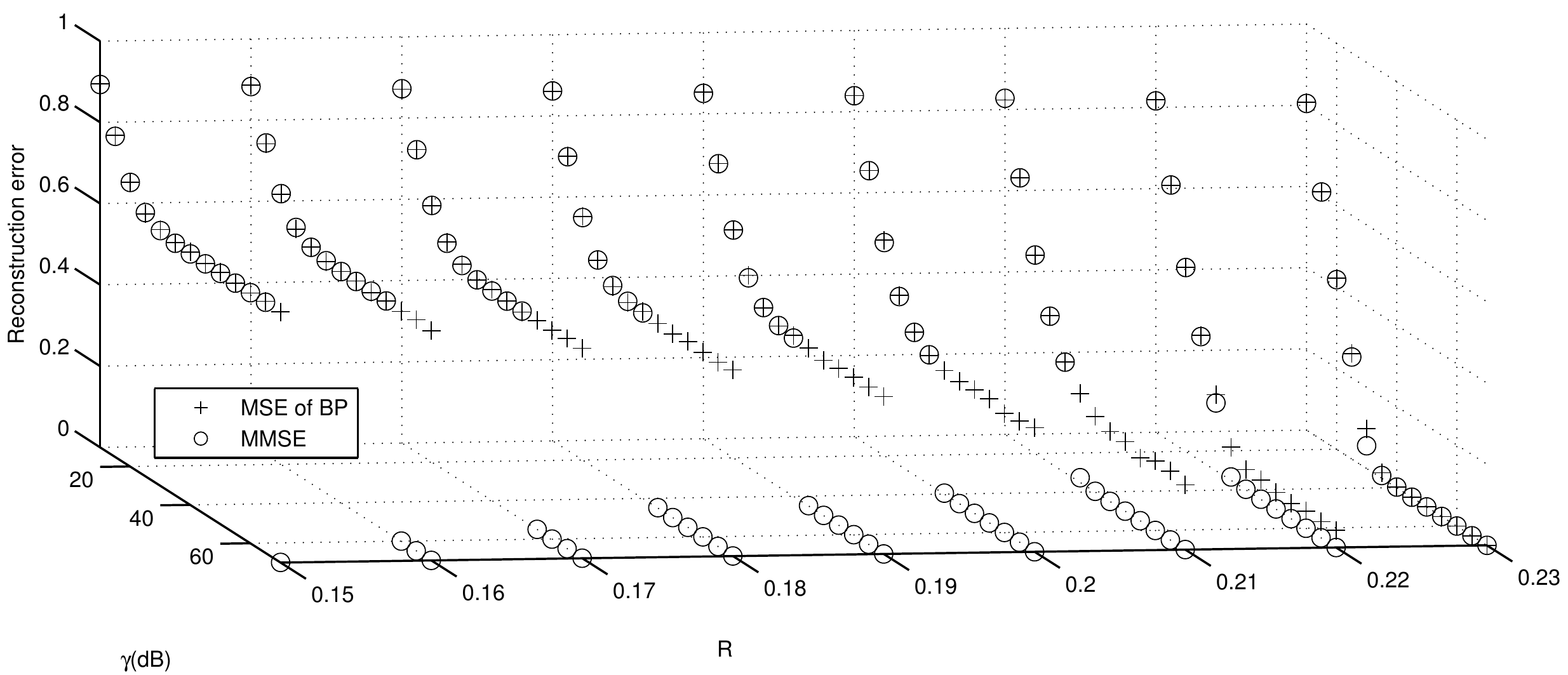}
\caption{Comparison of the MSE performance of BP to the theoretically optimal MMSE ($p=0.1$).}\label{fig.BP}
\end{figure*}

We used a BP solver, GAMP \cite{RanganGAMP2010}, to simulate a series of CS reconstruction problems. The cases where $R\in \{0.15,0.16,...,0.23\}$ and $\gamma\in \{10\text{dB},15\text{dB},..,70\text{dB}\}$ were simulated. For each $R$ and $\gamma$, 100 sparse Gaussian signals with a length of 10,000 and sparsity rate $p=0.1$, as well as the corresponding measurement matrices $\Phi$ and noise $z$, were generated. The matrices were i.i.d. Gaussian, with zero mean and unit norm rows; the noise followed $\mathcal{N}(0,1)$. We averaged the empirical MSE's for each of the 100 signals. The simulation results, as well as the theoretical MMSE given by Tanaka's equation, are illustrated in Figure~\ref{fig.BP}.

We can see that GAMP results in a reconstruction error that is close to the MMSE in the regions where there is only one fixed point, which includes Region~1 (corresponds to $R>0.21$ in Figure~\ref{fig.BP}) and Region~4 (corresponds to $\gamma$ roughly less than $40$ dB). However, in the region where there are several fixed points, GAMP operates at the smallest fixed point, as discussed by Guo and Wang \cite{Guo2006} and Rangan \cite{Rangan2010}. \emph{That is, BP is optimal in Regions~1, 2 and 4, while suboptimal in Region~3.} When $\gamma$ becomes large (for fixed $R$), we will inadvertently leave Region~2 and enter Region~3, where BP is suboptimal. The performance deficiency of BP is also the reason why we denote the boundary between Regions~1 and 3 by $R_{bp}(\gamma)$; it is the threshold above which BP is advantageous. Below the threshold, the MSE performance of BP is not satisfactory. For Region~3, which is part of the region having modest $\gamma$ and $R$, it might be possible to develop new algorithms that have better performance than BP.

\begin{figure}[t]
\centering
\includegraphics[width=9cm]{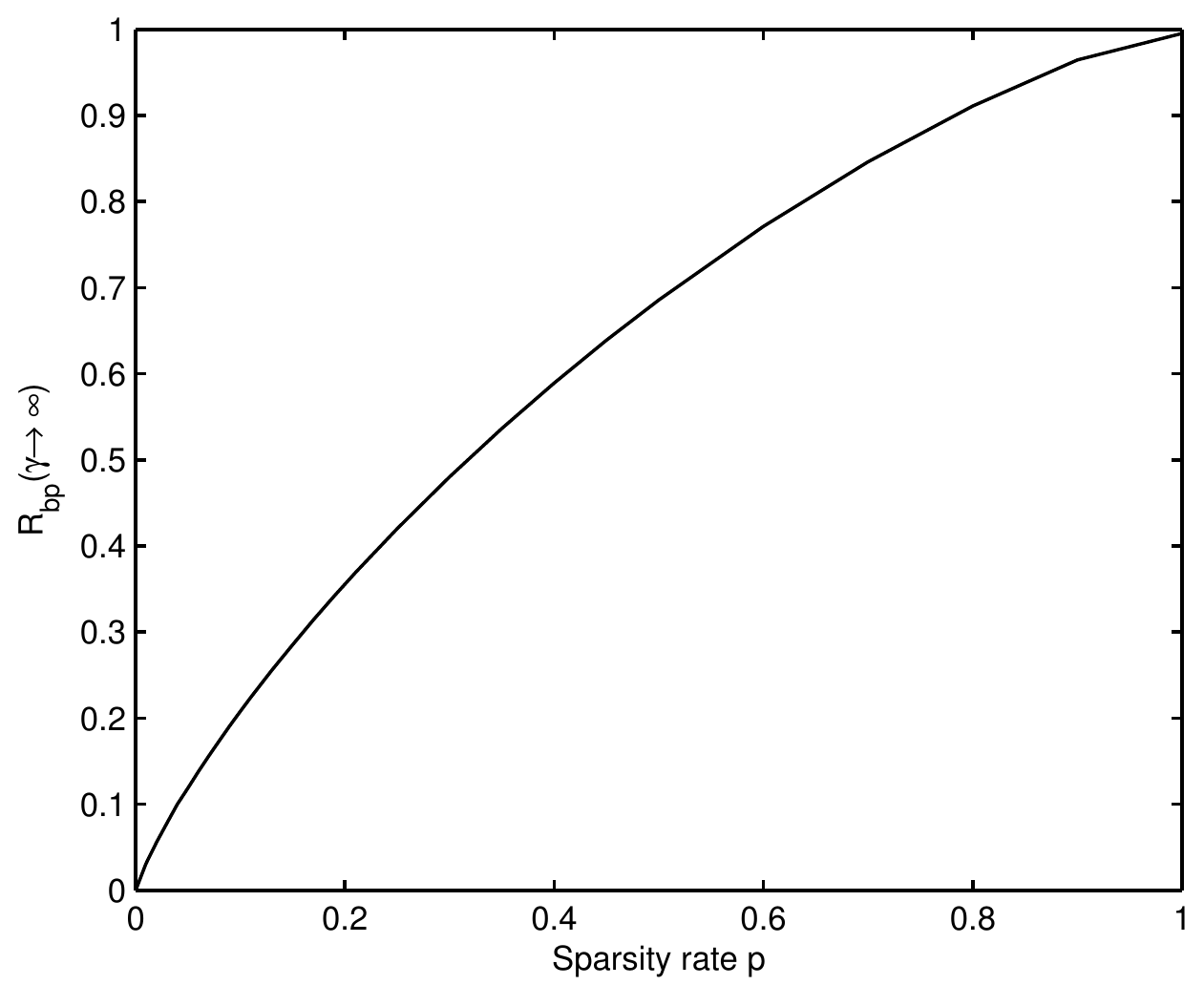}
\caption{Sparsity rate $p$ v.s. the threshold for BP to be advantageous}\label{fig.Rbp}
\end{figure}

From Figure~\ref{fig.SurfMMSE}, it appears that $R_{bp}(\gamma)$ converges to some $R_{bp}$ in the limit of large $\gamma$; more work is being done to demonstrate this convergence rigorously. We can approximate $R_{bp}$ for different $p$ by evaluating Tanaka's equation for some reasonably large $\gamma$. The result is shown in Figure~\ref{fig.Rbp}.

\section{Conclusion}
In this paper, we discussed the reconstruction performance of sparse Gaussian signals. By evaluating Tanaka's fixed point equation \eqref{eq.TE} and the free energy \eqref{eq.FreeEnergy}, we found that there are several regions where the reconstruction error behaves differently. We further compared the reconstruction performance of BP to the theoretically optimal performance, and showed that this state-of-the-art method has its limitations. Based on this observation, there is room for a new generation of reconstruction algorithms with better performance than BP in the region where both the inverse noise level and the measurement rate are modest.

\section*{Acknowledgment}
The authors thank Yihong Wu for his gracious help in enlightening us about his ground breaking work \cite{Wu2011,WuVerdu2011}. This work was supported by the National Science Foundation, Grant No. CCF-1217749, and by the U.S. Army Research Office, Grant No. W911NF-04-D-0003.

\bibliographystyle{IEEEtran}
\bibliography{IEEEabrv,cites}

\end{document}